\documentclass[prl, twocolumn, showpacs, superscriptaddress,longbibliography]{revtex4-1}
\usepackage{amsmath, amsfonts, amssymb, mathrsfs}
\usepackage{graphicx, subfigure, color}
\usepackage{bm}
\usepackage{braket}
\usepackage{epstopdf}
\usepackage[normalem]{ulem} 
\usepackage{color}
\usepackage[breaklinks]{hyperref}
\hypersetup{
   colorlinks,
   citecolor=blue,
   filecolor=blue,
   linkcolor=blue,
   urlcolor=blue
}
\graphicspath{Graphes}

\renewcommand{\underline}[1]{\uline{#1}}

\newcommand{\be}{\begin{equation}}
\newcommand{\ee}{\end{equation}}

\newcommand{\vectq}[4]{\begin{pmatrix} #1 \\ #2 \\ #3 \\ #4 \end{pmatrix}}

\newcommand{\ti}[1]{{\color{blue}{\underline{\textit{#1:}}}}}

\definecolor{orange}{rgb}{1,0.5,0}
\definecolor{grey}{rgb}{.6,.6,.6}

\definecolor{grey}{rgb}{.6,.6,.6}

\begin{document}

\title{Bipartite charge fluctuations in one-dimensional $\mathbb{Z}_2$ superconductors and insulators}

\author{Lo\"{i}c Herviou}
\affiliation{Centre de Physique Th\'{e}orique, \'{E}cole Polytechnique, CNRS, Universit\' e Paris-Saclay, 91128 Palaiseau, France}
\affiliation{Laboratoire Pierre Aigrain, \'Ecole Normale Sup\'erieure-PSL Research University, CNRS, Universit\'e Pierre et Marie Curie-Sorbonne Universit\'es, Universit\'e Paris Diderot-Sorbonne Paris Cit\'e, 24 rue Lhomond, 75231 Paris Cedex 05, France}
\author{Christophe Mora}
\affiliation{Laboratoire Pierre Aigrain, \'Ecole Normale Sup\'erieure-PSL Research University, CNRS, Universit\'e Pierre et Marie Curie-Sorbonne Universit\'es, Universit\'e Paris Diderot-Sorbonne Paris Cit\'e, 24 rue Lhomond, 75231 Paris Cedex 05, France} 
\author{Karyn~Le~Hur}
\affiliation{Centre de Physique Th\'{e}orique, \'{E}cole Polytechnique, CNRS, Universit\' e Paris-Saclay, 91128 Palaiseau, France}

\begin{abstract}
Bipartite charge fluctuations (BCF) have been introduced to provide an experimental indication of many-body entanglement. They have proved themselves to be a very efficient and useful tool to characterize quantum phase transitions in a variety of quantum models conserving the total number of particles (or magnetization for spin systems). In this Letter, we study the BCF in generic one-dimensional $\mathbb{Z}_2$ (topological) models including the Kitaev superconducting wire model, the Ising chain or various topological insulators such as the SSH model. The considered charge (either the fermionic number or the relative density) is no longer conserved, leading to macroscopic fluctuations of the number of particles. We demonstrate that at phase transitions characterized by a linear dispersion, the BCF probe the change in a winding number that allows one to pinpoint the transition and corresponds to the topological invariant for standard models. Additionally, we prove that a sub-dominant logarithmic contribution is still present at the exact critical point. Its quantized coefficient is universal and characterizes the critical model. Results are extended to the Rashba topological nanowires and to the XYZ model.

\end{abstract}

\date{\today}
\maketitle

\begin{figure}
\begin{center}
\includegraphics[width=0.9\linewidth]{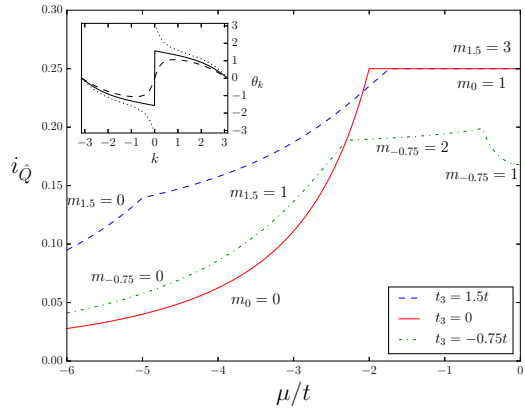}
\end{center}
\caption{(Main graph) Linear contribution to the BCF for the  extended Kitaev model as a function of the chemical potential. $m_{t_3}$ is the winding number of $\theta_k$, for $t_3=1.5t$, $t_3=0$ and $t_3=-0.75t$. $t_3=0$ corresponds to the Kitaev chain. On the other two lines, the system undergoes two transitions with changes in the winding numbers. The winding number here is also the number of Majorana edge states at each extremity\cite{Niu2012}. (Inset) $\theta_k$ as a function of $k$ in the Kitaev chain for $\mu=-3t$ (trivial phase, dashed line), $\mu=-t$ (topological phase, dotted line) and $\mu=-2t$ (QPT, continuous line). $\pi$ and $-\pi$ are identified for both $k$ and $\theta_k$. It exhibits a discontinuity precisely at the QPT. }
\label{fig:QFID1D}
\end{figure}

\ti{Introduction} Topological phases and topological quantum phase transitions (QPT) have become of tremendous importance in Condensed Matter physics during the last decade. These transitions, occurring at zero temperature, translate into significant change in the entanglement structure of the system and its ground states. Standard tools for analysing this entanglement are the von Neumann entanglement entropy (EE) and spectrum\cite{LiHaldane, Vidal2003, Kitaev2006, Levin2006, Thomale2009, Turner2011}. These entanglement measures detect the phase transitions and characterize some of the topological properties of the system. A common effect is for example the change in degeneracy of the entanglement spectrum, manifesting the appearance of topological zero-energy edge states when cutting the system\cite{LiHaldane, Turner2011}. In two-dimensional systems, a topological constant (that does not scale with the size of the considered subregion) also appears in the entanglement entropy\cite{Kitaev2006, Levin2006}. Despite some recent proposals and experimental efforts\cite{Islam2015}, these fundamentally theoretical quantities are challenging to experimentally measure, requiring copy of a quantum system and complex swap operations\cite{Cardy2011, Abanin2012, Pichler2016}.\newline
Alternative observables have been proposed to solve this conundrum, and the one this Letter focuses on is bipartite fluctuations, and more particularly \textit{bipartite charge fluctuations} (BCF)\cite{Klich2006, song2011entanglement, rachel2012detecting, song2012bipartite, Klich2014, Petrescu2014}. Let $A$ be a subregion of the total system $S$. We define the BCF as:
\begin{equation}
F_{\hat{Q}}(A) =  \langle(\sum\limits_{j\in A} \hat{Q}_j )^2\rangle-\langle\sum\limits_{j\in A} \hat{Q}_j \rangle^2=\langle(\sum\limits_{j\in A} \hat{Q}_j )^2\rangle_C,
\end{equation} 
where $\Braket{...}$ denotes the ground state average at zero temperature and $\hat{Q}_j$ is a local (unit-cell) charge operator (such as the electron number or the spin polarization). BCF have been introduced and studied in both one- and two-dimensional $U(1)$ models (\textit{i.e.} where the total charge is conserved), as a tool to detect and characterize quantum phase transitions and gapless phases (modes)\cite{rachel2012detecting, Petrescu2016}. In this context, they present strong similarities with EE, such as an area law for gapped ordered phases and a logarithmic growth for gapless (quasi-)ordered phases in one dimension. As an example of the latter, the study of BCF in Luttinger liquids gives a highly precise estimate for the Luttinger parameter\cite{Hsu2009}. BCF have also been used to characterize the superradiant transition in the Dicke model\cite{Nataf2012}. We also note recent works on bipartite fluctuations in spin chains, in relation with Many-Body Localization\cite{Singh2016}.  Experimentally, microwave cavities offer a flexible tool to probe charge polarization and BCF. Recent works\cite{Cottet2013,Dmytruk2015, Dmytruk2016} thus discuss how topological superconductors could be characterized under microwave radiation.\\
The aim of this Letter is to study the BCF in one-dimensional $\mathbb{Z}_2$ topological superconductors and insulators, where only the parity of the considered charge is conserved. For this family of models, the quantum phase transition is not described by a local order parameter, but by an abrupt change in a topological number. 
Let us first summarize our main results. Due to the non-conservation of the total charge in the system, long-range entanglement leads to a volume law for both gapped phases and critical points:
\begin{equation}\label{eq:fluctua}
F_{\hat{Q}}(A)= i_{\hat{Q}} l + b \log l + o(l),
\end{equation}
where $l$ is the size of the subregion $A$.  $i_{\hat{Q}}$ gives the fluctuations of the total charge in the system per unit length, also studied as the Quantum Fisher Information density\cite{Helstrom}. It vanishes for $U(1)$ systems with charge conservation. We explore $\mathbb{Z}_2$ transitions within a Bogoliubov framework by varying a control parameter (typically the chemical potential)  across a gapless point.  At the phase transition, $i_{\hat{Q}}$ presents a cusp corresponding to an abrupt change in the winding number of the Bogoliubov angle. The coefficient $b$ of the sub-leading log term in Eq.~\eqref{eq:fluctua} vanishes in gapped phases but not in gapless phases. Its value is even universal, independent of the microscopic parameters, when the gap closes at a single point in $k$-space. Details of the proofs and the various computations are presented in the Supplementary Materials\cite{SuppMat}.
\\
\ti{Models}
We consider in this work various realistic and interacting models to be addressed below. Let us start for simplicity with simple non-interacting models and consider a general Bogoliubov form in momentum space:
\begin{equation}
H = \frac{1}{2} \sum\limits_{k}  \Psi_k^\dagger(\varepsilon_k \sigma^z + \Delta_k \sigma^x) \Psi_k, \label{eq: TM1D}
\end{equation}
where $\varepsilon_k$ and $\Delta_k$ are continuous by part, and $\Psi_k$ is the Bogoliubov spinor. We are interested in models where $\varepsilon_k$ is even in momentum space while $\Delta_k$ is odd~\footnote{We have mostly in mind a quadratic (linear) behaviour for  $\varepsilon_k$ ($\Delta_k$) corresponding to a linear dispersion but the discussion also applies to higher powers of $k$.}.\\
Relevant topological superconducting models of this form include the Kitaev chain\cite{Kitaev2001}. $\varepsilon_k$ is then the kinetic energy, $\Delta_k$ a pairing term and $\Psi^\dagger_k$ takes the form $(c^\dagger_k, c_{-k})$. As a guide for the forthcoming discussion, we shall study an extended version of the Kitaev chain, where third-nearest-neighbor hopping and pairing have been added, leading to 
\begin{equation}\label{eq:dis}
\begin{split}
\varepsilon_k&=-\mu - 2t\cos(k)-2 t_3 \cos(3k),\\
 \Delta_k &= 2\Delta \sin(k) + 2\Delta_3 \sin(3k).
\end{split}
\end{equation}
Here, $\mu$ is the chemical potential, $t$ ($t_3$) is the (third-nearest neighbor) hopping and $\Delta$ ($\Delta_3$) the (third-nearest neighbor) hopping and the lattice spacing has been fixed to unity. Without loss of generality we choose $t=\Delta$ and $t_3=\Delta_3$. This model presents a richer phase diagram with up to $3$ Majorana fermions at each extremity \cite{Niu2012}, behaving as if the system had up to $3$ different bands (the extension to $M$ end states is straightforward). This model enables us to study transitions between phases with $0$ and $1$ Majorana end states, but also between $0$ and $2$ ($0$ and $3$) where the gap closes at two (three) different momenta.\\
The local charge operator can be written as:
\begin{equation}
\hat{Q}_j = \frac{q_e}{2} \Psi^\dagger_j \sigma^z \Psi_j,\label{eq:Charge}
\end{equation}
where $q_e$ is the charge by unit cell (here $q_e=1$).\\
Additionally, topological insulators can be described by the same formalism. We present here two of such models. The first one is a typical model of topological insulator \cite{Guo2011, Shen}:
\begin{multline}
H_{I, 1}=-\mu\sum\limits_j  c^\dagger_{j} \sigma^z c_j - t\sum\limits_j (c^\dagger_{j+1} \sigma^z c_{j}+h.c.)\\
+\sum\limits_j i \Delta (c^\dagger_{j} \sigma^x c_{j+1}-h.c.),
\end{multline}
where $c_j$ are spin-$\frac{1}{2}$ fermionic annihilation operators, $t$ and $\Delta$ are orthogonal Rashba spin-orbit couplings and $\mu$ is a Zeeman field. In momentum space, $H_{I, 1}$ takes exactly the form given in Eq. \ref{eq: TM1D}, with $\Psi^\dagger_k = (c^\dagger_{k, \uparrow}, c^\dagger_{k, \downarrow})$. The considered charge (the spin polarization) is also given by Eq. \ref{eq:Charge}, but with two charges  ($q_e=2$) per unit cell. The second one is the Su-Schrieffer-Heeger model\cite{SSHmodel}:
\begin{equation}
H_{I, 2}=- t_1\sum\limits_j(c^\dagger_{j, A} c_{j, B}+h.c.)-t \sum\limits_j(c^\dagger_{j, B} c_{j+1,A}+h.c.),
\end{equation}
where $c_{j, A/B}$ describe two different species of fermions. Identical formalism and results are recovered by taking $(\sigma^z, \sigma^x) \rightarrow (\sigma^x, \sigma^y)$ in Eq. \ref{eq: TM1D} and \ref{eq:Charge}, with $\Psi_k^\dagger=(c^\dagger_{k,A}, c^\dagger_{k, B})$ and $q_e=2$. $t_1/2$ plays the role of Kitaev's chemical potential.\\
Finally, spin chains such as the Quantum Ising model also follow this formalism after a  Jordan-Wigner transform\cite{Subirbook}. The charge then corresponds to the transverse polarization.\\
For all these models, the energy spectrum is given by $\pm \sqrt{\varepsilon_k^2+\Delta_k^2}$, and consequently both $\varepsilon_k$ and $\Delta_k$ need to vanish at a QPT. We introduce the Bogoliubov angle:
\begin{equation}
\theta_k = \text{Arg}(\varepsilon_k-i\Delta_k).\label{eq:theta1D}
\end{equation} 
The winding number of $\theta_k$, \begin{equation} m= \oint \frac{d\theta_k}{2\pi},\end{equation} where $k$ is summed over the Brillouin zone (BZ), probes the family of QPT studied in this part. For the previous fermionic models, it actually corresponds to the topological index\cite{Tewari2011, Trif2012}, and the number of edge states with open boundary conditions (Majorana fermions for superconductors, complex fermions for insulators). It has been experimentally measured in photonic and Cold Atoms setups\cite{Kitagawa2012, Atala2013, Mugel2016, Flurin2016}. Interestingly, the angle $\theta_k$ is ill-defined right at the QPT when both $\varepsilon_k$ and $\Delta_k$ vanish and a discontinuity occurs in $\theta_k$ on the condition that $\varepsilon_k$ vanishes faster than  $\Delta_k$ (as shown in the inset of Fig. \ref{fig:QFID1D}). This discontinuous behaviour is in fact related to the abrupt change in the winding number $m$ across the transition. As demonstrated below, it is also responsible for singularities in the BCF. 
\\
\ti{Generalities} from Wick theorem, the BCF for a linear subregion $A$ of size $l$ can be computed in the thermodynamic limit: 
\begin{multline}
F_{\hat{Q}}(A) = q_e l \iint_{\text{BZ}^2} \frac{dk dq}{16\pi^2} f(k-q, l) \\
\left( 1-\cos(\theta_k) \cos(\theta_q)+\sin(\theta_k) \sin(\theta_q) \right), \label{eq:BCF1D}
\end{multline}
where the integration carries on the Brillouin Zone (BZ) and $f(k, l)$ is the Fej\'{e}r Kernel:
\begin{align}
f(k,l)&=\frac{\sin^2(\frac{k l}{2})}{l \sin^2(\frac{k}{2})}=\sum\limits_{j=-l}^l (1-\frac{|j|}{l})e^{i (j k)}\\
f(k, l) &\rightarrow 2\pi \delta(k), \qquad  \text{when $l\rightarrow +\infty$.} \label{eq:FejDirac}
\end{align}
The BCF can be re-written in the convenient form:
\begin{multline*}
F_{\hat{Q}}(A) = q_e (\frac{l}{4} + \frac{1}{4} \sum\limits_{j=-l}^l (l-|j|)\\
 (|\mathcal{FT}\{\sin(\theta_k)\}(j)|^2-|\mathcal{FT}\{\cos(\theta_k)\}(j)|^2)),
\end{multline*}
where $\mathcal{FT}\{f(\theta_k)\}$ is the Fourier transform of $f(\theta_k)$. At large $j$, continuity of $\theta_k$ in gapped phases translates into $|\mathcal{FT}\{f(\theta_k)\}(j)|=\mathcal{O}(\frac{1}{j^2})$, while the discontinuities at the QPT give a different scaling $|\mathcal{FT}\{f(\theta_k)\}(j)|=\mathcal{O}(\frac{1}{j})$. Basic series analysis leads to the general form for the BCF:
\begin{equation}
F_{\hat{Q}}(A)= i_{\hat{Q}} l + b \log(l) + \mathcal{O}(1), \label{eq:GenForm}
\end{equation}
where $b$ is non-zero only at the QPT. \newline
\ti{Linear contributions}  from Eq. \ref{eq:BCF1D} and   \ref{eq:FejDirac}, $i_{\hat{Q}}$ is the density of charge fluctuations in the total system.
\begin{align}
i_{\hat{Q}} &= \lim\limits_{L\rightarrow +\infty} \frac{1}{L}\Braket{\hat{Q}^2}_C=  q_e \int\limits_{\text{BZ}} \frac{dk}{4\pi} \sin^2(\theta_k),
\end{align}
where $L$ is the total length of the system and $\hat{Q}$ the total charge. Remarkably, these fluctuations coincide at $T=0$ with the Quantum Fisher Information density\cite{Helstrom} (QFID) associated to the chemical potential. The Quantum Fisher Information has been used to characterize several transitions\cite{Wang2014,  Hauke2015, Wu2016, Ye2016, Monroe2016} or study quenches in the quantum Ising model\cite{Pappalardi2017}, and gives a bound on the precision with which one can evaluate the chemical potential. Additionally, for superconducting models, the QFID allows for a direct evaluation of the superconducting gap: for Kitaev chain for example, $i_{\hat{Q}}=\frac{|\Delta|}{2(|\Delta|+2|t|)}$ in the topological phase\cite{Herviou1}. Other noise measurements (fluctuations) have been suggested to measure $\Delta$\cite{Altman2004}. \\
Figure \ref{fig:QFID1D} represents $i_{\hat{Q}}$ as a function of the chemical potential for both the Kitaev chain and its extended version\cite{Niu2012}.  $i_{\hat{Q}}$ is continuous but presents a cusp at transitions where the winding number of $\theta_k$ changes. For systems in which the winding number is a topological invariant, our analysis shows that $i_{\hat{Q}}$ reveals the topological nature and location of the transition. \\
\ti{Logarithmic contribution} The logarithmic term appears if there are discontinuities in $\theta_k$ at points in $k$-space where the gap closes. For a  closure at a single momentum $k_0$, $b$ directly measures this discontinuity,
\begin{equation}
b=q_e\frac{\left( \cos(\theta_{k_0^+})-\cos(\theta_{k_0^-})\right)^2-\left( \sin(\theta_{k_0^+})-\sin(\theta_{k_0^-})\right)^2}{2\pi^2}.
\end{equation}
The assumed even (odd) symmetry of $\varepsilon_k$ ($\Delta_k$) imposes a gap single-closure at either $k_0=0$ or $k_0=\pi$. Hence, $\theta_{k_0^\pm} = \pm \pi/2$ and one obtains a \textit{universal} coefficient for the logarithmic fluctuations:
\begin{equation}\label{eq:b}
b=-\frac{q_e}{2\pi^2}
\end{equation}
in agreement with Conformal Field Theory. The negative sign reveals the existence of a Majorana mode, as opposed to a $U(1)$ Luttinger model\cite{Herviou1}. The factor $2$ difference ($q_e=2$ vs $q_e=1$) between the 1D (topological) insulator models and the superconductor Kitaev chain originates from the doubling in the number of degrees of freedom. The corresponding critical theories involves complex (Majorana) fermions for insulators (superconductors) with the respective central charges $c=1$ and $1/2$. The same factor $2$ is also found by comparing the degeneracies due to the edge states, $2$ for the standard ($t_3=0$) Kitaev chain and $4$ for the SSH model.

In some cases, the QPT is characterized by a gap closing at multiple momenta. A simple example is provided with $t=0$ and $\mu=-2t_3$ in Eq.~\eqref{eq:dis}. The gap closes at $k=0$ and $k=\pm \frac{2 \pi}{3}$ leading to $b=-\frac{3}{2\pi^2}$. In the general case, only a bound for $b$ can be derived,
\begin{equation}
|b|\leq \frac{q_e N}{2\pi^2},
\end{equation}
 when the gap closes $N$ times. This bound also applies for free fermions with $U(1)$ charge conservation where the gap closes twice at  $k = \pm k_F$ and the BCF are given by  $\frac{1}{\pi^2} \log(l) + \mathcal{O}(1)$. \newline
The structure of the QPT and universality can be further examined by looking at the structure factor of the BCF. We define:
\begin{align}
SF_{\hat{Q}}(A, \phi)&=\langle|\sum\limits_{j \in A} e^{i \phi j} \hat{Q}_j|^2\rangle-|\langle\sum\limits_{j \in A} e^{i \phi j} \hat{Q}_j\rangle|^2\\
&=\langle|\mathcal{FT_A}\{\hat{Q} \}(\phi)|^2\rangle - |\langle\mathcal{FT_A}\{\hat{Q} \}(\phi) \rangle|^2,
\end{align}
where $\mathcal{FT_A}\{\hat{Q} \}$ is the Fourier transform of the charge in the region A. The integral form is similar to the one in Eq. \ref{eq:BCF1D}, but where $f(k-q,l)$ is replaced by $f(k+\phi-q, l)$. The additional phase leads to oscillations that generally destroy the logarithmic contributions so that $b(\phi)$ vanishes. Nevertheless, at some definite values of the phase $\phi=k_{i}-k_{j}$ matching the difference between two momenta $k_i, k_j$ at which the gap closes, logarithmic contributions to $SF_{\hat{Q}}(A, \phi)$ reemerge and $b(\phi)$ takes a finite value. For instance, assuming that the gap closes twice at $k = \pm k_1$, then we obtain the universal value $b(2 k_1) = -\frac{q_e}{2\pi^2}$. Similarly, discontinuities of $\partial_\phi i_{\hat{Q}}(\phi)$ occur at the same values.

\begin{figure}
\begin{center}
\includegraphics[width=0.9\linewidth]{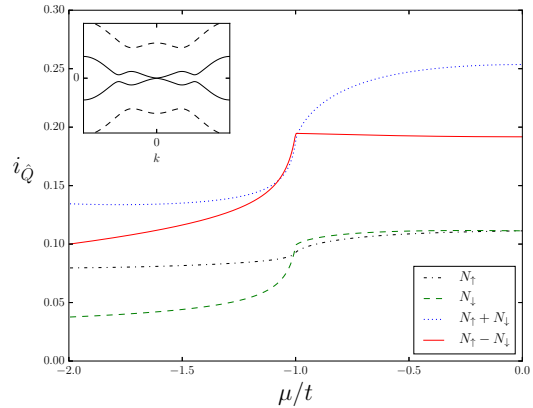}
\end{center}
\caption{QFID for the Rashba model of topological superconductors across the topological QPT. Parameters are $\Delta=t$, $\lambda=0.2t$ and $V_z=-\sqrt{2}t$. The transition is here driven by the chemical potential $\mu$ and occurs at $\mu=-t$. Discontinuity of the derivative of the QFID can be seen for example in the spin BCF. In inset, the band structure at the QPT where the gap vanishes at $k=0$. Each band is a mixture of the two spin polarizations.}
\label{fig:Rashba}
\end{figure}

\textit{Finite-size corrections:} in order to analyze data in simulations or real systems, one needs to take into account finite-size effects. We have computed numerically (and analytically for $t=\Delta$) the BCF for a finite system for the Kitaev chain. An additional finite size correction compared to the BCF in charge-conserving system appear:
\begin{multline}
F_{\hat{Q}}(l) = \frac{|\Delta| l}{2|\Delta| +2|t|} -\frac{1}{2\pi^2} \log(\frac{L}{\pi}\sin(\frac{l \pi}{L}))-\frac{l^2}{4L^2}  + \mathcal{O}(1)
\end{multline}
\ti{Extensions} The results so far have been derived for the generic toy model described in Eq. \ref{eq: TM1D}, but can be extended to more realistic systems. We have analytically computed the BCF in the Rashba nanowire model\cite{wire1, wire2} for topological superconductors where the wire hosts spin-$\frac{1}{2}$ fermions with strong spin-orbit coupling. The superconducting proximity effect implies that charge is not conserved but the total parity is. The results of our calculations are displayed in Figure \ref{fig:Rashba}. They show that clear cusps in the QFID of all spin combinations probe the topological phase transition with high precision. These features can also be related to the abrupt change in the topological winding number characterizing the nanowire. Similarily, a quantized logarithmic term also develops in the bipartite spin fluctuations at the QPT with the same constant $b$ as Eq.~\eqref{eq:b}. \newline
We finally explore the effect of adding nearest-neighbour interaction to the Kitaev chain model, corresponding to a XYZ spin chain\cite{Sela2011, Hassler2012}, with the aim of generalizing our discussion to interacting systems. Our MPS computations using the ALPS library\cite{Bauer2011, dolfi2014matrix} demonstrate that the BCF singularities at the QPT are not qualitatively  modified by interactions as long as these interactions preserve the nature of the QPT. Hence, the linear term in the BCF exhibits a cusp at the QPT, precisely pinpointing its exact location, and the sub-leading logarithmic term emerges right at the gapless QPT with a coefficient depending on the strength of interactions. \\
Details of the computation for the Rashba nanowires and examples of the numerical results are exposed in the Supplementary Materials\cite{SuppMat}.

\ti{Conclusions} we have shown that the bipartite fluctuations of the charge or spin characterize phase transitions in $\mathbb{Z}_2$ topological systems. The scaling analysis of these fluctuations with the length of the sub-region reveals quantum phase transitions in two ways: the leading linear behaviour exhibits a cusp and a subleading quantized logarithmic term appears at the transition. For free electrons, an exact relation has been drawn between the full-counting statistics associated with the partial charge $Q_A$ and the entanglement and Renyi entropies; extending these relations for superconductors remains an open question\cite{Klich2009, song2012bipartite, Klich2014}. Extension of these results to higher dimensional topological systems will be the subject of a following paper\cite{Herviou5}. In particular, in two-dimensional systems such as the $p+ip$ superconductors, we find a divergence of the second derivative of the QFID and a quantized logarithmic scaling at the topological phase transition.

\ti{Acknowledgement} This work has benefited from useful discussions with W. Witczak-Krempa and A. Mesaros. We acknowledge financial support from the PALM Labex, Paris-Saclay, Grant No. ANR-10-LABX-0039 and from the German Science Foundation (DFG) FOR2414.

\bibliography{Fluctuations2}

\clearpage

\setcounter{equation}{0}
\setcounter{figure}{0}
\setcounter{table}{0}

\renewcommand{\theequation}{S\arabic{equation}}
\renewcommand{\thefigure}{S\arabic{figure}}
\renewcommand{\bibnumfmt}[1]{[S#1]}
\renewcommand{\citenumfont}[1]{S#1}

\onecolumngrid
\begin{center}
\large{Supplementary Materials: Bipartite charge fluctuations in one-dimensional $\mathbb{Z}_2$ superconductors and insulators.}
\end{center}

In these Supplementary Materials, we provide for reference some details on the Bogoliubov formalism adopted to describe the different non-interacting topological superconductors and insulators, including a detailed diagonalization of the Rashba superconductors. Proof of the discontinuity of the derivative of the Quantum Fisher information at a topological phase transition is sketched. Finally, we prove the appearance of secondary logarithmic contributions and discontinuities in the structure factor of the bipartite charge fluctuations.\\\\
\twocolumngrid

\section{Bogoliubov formalism}
In this section, we detail the Bogoliubov formalism that allow to exactly diagonalize the Kitaev chain[\hyperref[S1]{S1}] and its extended version with periodic boundary conditions. We also mention sketch the computation for a typical model of topological insulator[\hyperref[S1]{S2}] and the SSH chain[\hyperref[S1]{S3}].
\subsection{Bogoliubov quasi-particles formalism for the extended Kitaev model}
Let us first introduce the real space Hamiltonian for a Kitaev chain with additional third nearest-neighbor hopping and pairing term[\hyperref[S1]{S4}].
\begin{multline}
H_{K,3} =  -t \sum\limits_j( c^\dagger_j c_{j+1}+c^\dagger_{j+1} c_j + \frac{t_3}{t}(c^\dagger_j c_{j+3} + c_{j+3}^\dagger c_j)) \\
+\Delta \sum\limits_j( c^\dagger_j c^\dagger_{j+1}+c_{j+1} c_j + \frac{\Delta_3}{\Delta}(c^\dagger_j c^\dagger_{j+3} + c_{j+3} c_j)) \\
-\mu \sum\limits_j c^\dagger_j c_j,
\end{multline}
where $\mu$ is the chemical potential, $t$ ($t_3$) the (third-) nearest neighbor hopping and $\Delta$ ($\Delta_3$) a (third-) nearest neighbor pairing term. $c_j$ is the fermionic annihilator operator at site $j$. We consider periodic boundary conditions such that the quadratic Hamiltonian can be easily diagonalized in momentum space. \\
The Fourier transform convention we use is : $c_k = \frac{e^{-i \frac{\pi}{4}}}{\sqrt{L}}\sum\limits_{j=1}^L e^{-ikj}c_j$, where $L$ is the total size of the system. We define $\Psi_{k}^\dagger=(c^\dagger_k, c_{-k})$. Forgetting constant terms, we can therefore write the Hamiltonian as :
\begin{equation}H_{K,3}=\frac{1}{2} \sum\limits_{k} \Psi_{k}^\dagger h(k) \Psi_{k}, \label{eq:Ham}\end{equation}
with 
$$h(k) = \begin{pmatrix}
\varepsilon_k & \Delta_k \\
\Delta_k & - \varepsilon_k
\end{pmatrix},$$
where
\begin{align*}
\varepsilon_k&=-\mu - 2t\cos(k)-2 t_3 \cos(3k),\\
 \Delta_k &= 2\Delta \sin(k) + 2\Delta_3 \sin(3k),
\end{align*} 

\begin{figure}
\begin{center}
\includegraphics[width=0.9\linewidth]{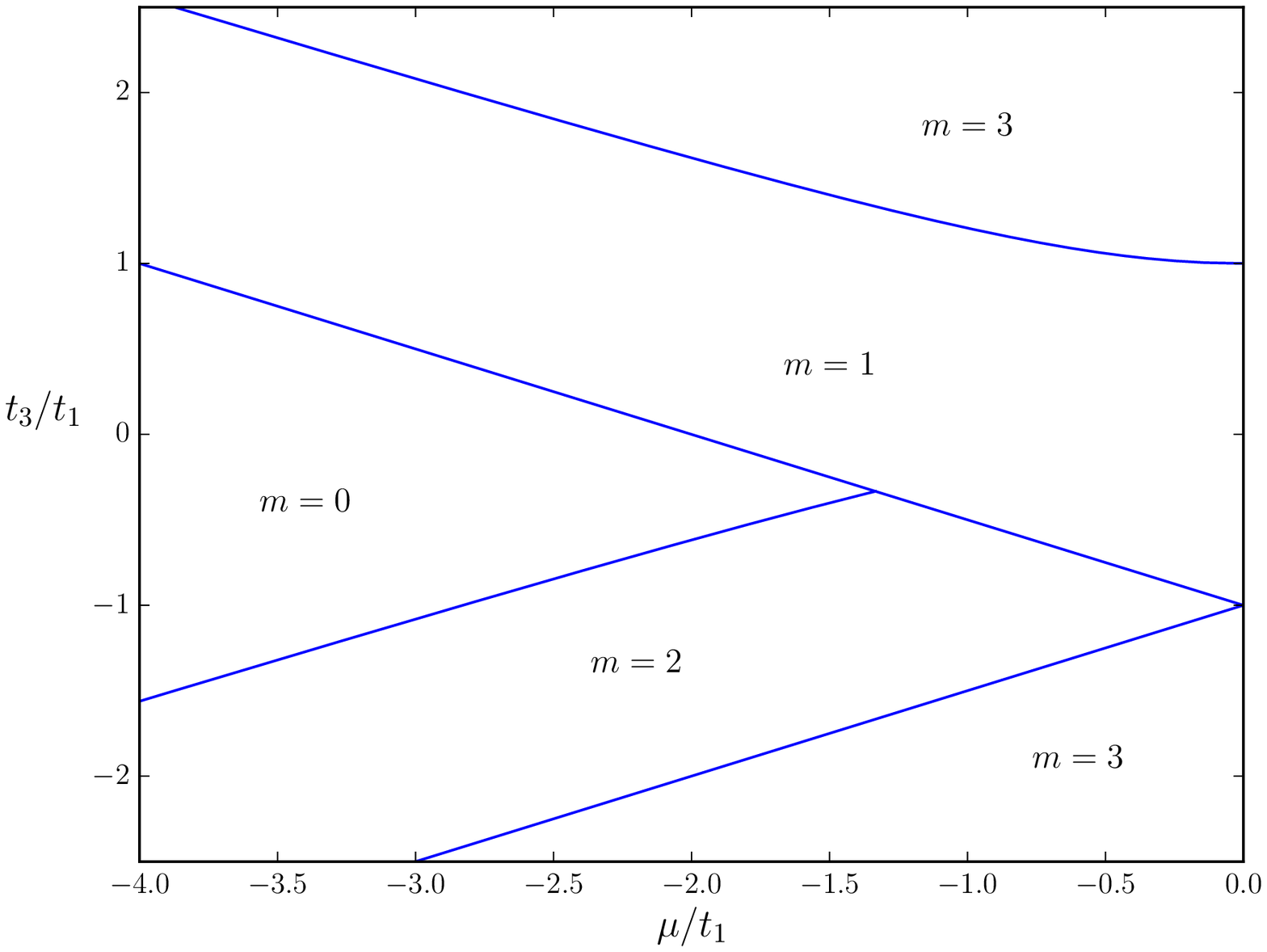}
\end{center}
\caption{Exact phase diagram for the extended Kitaev model with third-nearest neighbor hopping and pairing, with $t=\Delta$ and $t_3=\Delta_3$. $m$ is the winding number of $\theta_k$. It corresponds to the number of Majorana edge modes that would appear if the system had open boundaries. The diagram is symmetric under particle-hole transformation ($\mu \rightarrow -\mu$).}
\label{fig:PDKE}
\end{figure}

We define the angle $\theta_k$ by $\theta_k = \text{Arg}(\epsilon_k-i\Delta_k)$, such that:
$$h(k) = \sqrt{\varepsilon_k^2+\Delta_k^2}\begin{pmatrix}
\cos \theta_k & -\sin \theta_k \\
-\sin \theta_k & - \cos \theta_k 
\end{pmatrix},$$
We introduce the Bogoliubov quasi-particle operators $\eta_{k} = \cos(\theta_k/2) c_k - \sin(\theta_k/2) c_{-k}^\dagger$ that diagonalize the Hamiltonian and $E_k=\sqrt{\varepsilon_k^2+\Delta_k^2}$.
$$H_{K,3}=\sum\limits_{k} E_k \eta^\dagger_{k} \eta_{k}.$$
The ground state is simply the vacuum state for the $\eta$ operators $\Ket{0}_\eta$. A phase transition occurs when $E_k$ vanishes. To simplify the discussion, we fix $t=\Delta$ and $t_3=\Delta_3$ in the rest of the Section. Figure \ref{fig:PDKE} presents the exact phase diagram for the extended Kitaev model. In a system with open boundary condition, the model can present up to $m=3$ edge states. The winding of $\theta_k$ defined by \begin{equation}
\oint \frac{d \theta_k}{2 \pi}
\end{equation}
is a good topological invariant. It actually counts (up to a sign) the number of Majorana edge states and characterize the different phases and phase transitions.[\hyperref[S1]{S5}, \hyperref[S1]{S6}]\\
The charge we are interested in is defined by:
\begin{equation}
\hat{Q}_j = \frac{1}{2} \Psi^\dagger_j \sigma^z \Psi_j ,\label{eq:ChargeSupra}
\end{equation}
where  $\Psi_j$ the real-space version of the spinor $\Psi_k$. It corresponds to the fermion number (up to a constant).

\subsection*{Typical topological insulator}
We also study a typical example of topological insulator[\hyperref[S1]{S2}] described by the Hamiltonian:
\begin{equation}
H_{I, 1}=-\mu c^\dagger_{j} \sigma^z c_j - t (c^\dagger_{j+1} \sigma^z c_{j}+h.c.)+ i \Delta (c^\dagger_{j} \sigma^x c_{j+1}-h.c.),
\end{equation}
where $c$ are spin-$\frac{1}{2}$ fermionic annihilation operators, $t$ and $\Delta$ are orthogonal Rashba spin-orbit couplings and $\mu$ is a Zeeman field. If we define the spinor $\Psi^\dagger_k = (c^\dagger_{k, \uparrow}, c^\dagger_{k, \downarrow})$, the Hamiltonian in momentum space is given by Eq. \ref{eq:Ham}, with $t_3=\Delta_3=0$. Edge states are now complex fermions. The considered charge is the spin polarization:
\begin{equation}
\hat{Q}_j =  \Psi^\dagger_j \sigma^z \Psi_j = c^\dagger_{j \uparrow} c_{j, \uparrow}-c^\dagger_{j \downarrow} c_{j, \downarrow}
\end{equation}
\subsection{Su-Schrieffer-Heeger model}
The Su-Schrieffer-Heeger (SSH) model[\hyperref[S1]{S3}] is a simple model of dimerized fermions. It is a topological insulator model. Its Hamiltonian can be written as:
\begin{equation}
H_{I, 2}=- t_1(c^\dagger_{j, A} c_{j, B}+h.c.)-t (c^\dagger_{j, B} c_{j+1,A}+h.c.),
\end{equation}
Let $\Psi_k^\dagger = (c^\dagger_{k, A}, c^\dagger_{k, B})$. In momentum space, the Hamiltonian can be rewritten as:
\begin{equation}
H_{I,2}=\frac{1}{2} \sum\limits_{k} \Psi_{k}^\dagger( -2(t_1+t\cos(k))\sigma^x -2 t \sin(k) \sigma^y) \Psi_{k}
\end{equation}
Using similar conventions as for superconductors, we define $\varepsilon_k =  -2(t_1+t\cos(k))$, $\Delta_k=-2t \sin(k)$ and $\theta_k = \text{Arg}(\varepsilon_k - i\Delta_k)$. The Bogoliubov operators diagonalizing $H_{I,2}$ are $\eta_{k, \pm}^{SSH} = \frac{e^{\mp i \frac{\theta_k}{2}}}{\sqrt{2}} c_{k,A} \pm \frac{e^{\pm i \frac{\theta_k}{2}}}{\sqrt{2}} c_{k,B}$. A phase transition between two distinct topological phases occurs when $t_1=t$, where one recovers a chain of free fermions. The winding number of $\theta_k$ is a good topological number and still counts the number of edge states. These edge states are also complex fermions.
The charge is defined by:
\begin{equation}
\hat{Q}_j = \Psi^\dagger_j \sigma^x \Psi_j=c^\dagger_{j,A} c_{j,B} + c^\dagger_{j, B} c_{j,A}.\label{eq:ChargeSSH}
\end{equation}

\section{Bipartite charge fluctuations in simple models}
In this section, we present some details on the computation of the fluctuation in the various non-interacting models considered in the main text.
\subsection{Details for superconductors and quantum XY chain}
The bipartite charge fluctuations are given by:
\begin{align}
F_{\hat{Q}}(A) &=  \langle(\sum\limits_{j\in A} \hat{Q}_j )^2\rangle-\langle\sum\limits_{j\in A} \hat{Q}_j \rangle^2=\langle(\sum\limits_{j\in A} \hat{Q}_j )^2\rangle_C\\
&=\frac{1}{L^2}\sum\limits_{j_1, j_2 = 1}^l \sum\limits_{k, k', q, q'} e^{i (k'-k)j_1} e^{i (q'-q)j_2}\langle c^\dagger_{k} c_{k'} c^\dagger_q c_{q'} \rangle_C
\end{align} 
Rewriting the original operators as a function of the $\eta_k$, we can deduce the following average in the ground state:
\begin{align*}
\Braket{c^\dagger_k c_q}&= \delta_{q,k}\sin(\theta_k/2)^2\\
\Braket{c^\dagger_k c^\dagger_q} &= \frac{\delta_{q, -k}}{2} \sin(\theta_k).
\end{align*}
Then using Wick theorem, one can compute the 4-fermions correlators. With the relation:
\begin{equation}
|\sum\limits_{j_1, j_2 = 1}^l e^{i (k-q) (j_1-j_2)}|^2=\frac{\sin^2(\frac{(k-q) l}{2})}{ \sin^2(\frac{k-q}{2})},
 \end{equation}
 it is straightforward to obtain: 
\begin{multline}
F_{\hat{Q}}(A) = l \iint_{\text{BZ}^2} \frac{dk dq}{16\pi^2} f(k-q, l) \\
\left( 1-\cos(\theta_k) \cos(\theta_q)+\sin(\theta_k) \sin(\theta_q) \right), \label{eq:BCF1D-supra}
\end{multline}
where the integration carries on the whole Brillouin Zone (BZ) and $f(k, l)$ is the Fej\'{e}r Kernel:
\begin{align}
f(k,l)&=\frac{\sin^2(\frac{k l}{2})}{l \sin^2(\frac{k}{2})}=\sum\limits_{j=-l}^l (1-\frac{|j|}{l})e^{i (j k)}\\
f(k, l) &\rightarrow 2\pi \delta(k), \qquad  \text{when $l\rightarrow +\infty$.} \label{eq:FejDiracSM}
\end{align}
\subsection{Computation for the insulators}
Let us focus on the spin-orbit model. The BCF are now given by:
\begin{align}
F_{\hat{Q}}(A) &=  \sum\limits_{j_1, j_2=1}^l \langle(c^\dagger_{j_1, \uparrow} c_{j_1, \uparrow}- c^\dagger_{j_1, \downarrow} c_{j_1, \downarrow}) (c^\dagger_{j_2, \uparrow} c_{j_2, \uparrow}- c^\dagger_{j_2, \downarrow} c_{j_2, \downarrow})  \rangle_C
\end{align}
While $4$ different correlators appear, anomalous correlators such as $\langle c^\dagger c^\dagger \rangle$ vanish, leading to the general formula:
\begin{multline}
F_{\hat{Q}}(A) = q_e l \iint_{\text{BZ}^2} \frac{dk dq}{16\pi^2} f(k-q, l) \\
\left( 1-\cos(\theta_k) \cos(\theta_q)+\sin(\theta_k) \sin(\theta_q) \right), \label{eq:BCF1D-uni}
\end{multline}
For the SSH model, the computation is very similar, and while the two-fermions correlators differ, the final result is identical.\\
We note the convenient form:
\begin{multline}
F_{\hat{Q}}(A) = q_e (\frac{l}{4} + \frac{1}{4} \sum\limits_{j=-l}^l (l-|j|)\\
 (|\mathcal{FT}\{\sin(\theta_k)\}(j)|^2-|\mathcal{FT}\{\cos(\theta_k)\}(j)|^2)),\label{eq:BCF1D-fourier}
\end{multline}
where $\mathcal{FT}\{f(\theta_k)\}$ is the Fourier transform of $f(\theta_k)$.

\subsection{Discontinuity  of $i_{\hat{Q}}$}
In this part, we prove the discontinuity of the derivative of $i_{\hat{Q}}$ at a simple phase transition. We first assume a QPT where the gap closes only at $k=0$. We consider transitions driven by an effective chemical potential such that $\varepsilon_k = \delta\mu + \delta \varepsilon_k$, with the transition occuring for $\delta \mu =0$, and $\delta\varepsilon_k \propto k^2$ when $k\ll 1$. As we consider linear spectrum at the phase transition, $\Delta_k \propto k$ when $k\ll 1$.\\
This transition leads to a change in the winding number of $\theta_k$, as $\theta_0 =0$ for $\delta \mu >0$ and $\theta_0 = \pi$ if $\delta \mu <0$ (assuming that the $\theta_k$ is only slightly changed by $\delta \mu$ far from $k=0$).\\
From F\'{e}jer Kernel's properties, one obtain:
\begin{equation}
i_{\hat{Q}} = q_e\int\limits_0^{2\pi} \sin^2(\theta_k) \frac{dk}{4 \pi}
\end{equation}
We define $DI(x) = \partial_{\delta \mu} i_{\hat{Q}} |_{\delta \mu = x}$. Let us compute the difference between the derivative on two sides of the transition. Let $x>0$ and $\Delta DI(x)= \frac{2\pi}{q_e} (DI(x)-DI(-x))$:
\begin{align}
\Delta DI(x) &=\int\limits_0^{2\pi} \Delta^2_k ( \frac{x+\delta \varepsilon_k}{((x+\delta \varepsilon_k)^2+\Delta^2_k)^2} \nonumber \\
&\qquad \qquad -\frac{-x+\delta \varepsilon_k}{((-x+\delta \varepsilon_k)^2+\Delta^2_k)^2}) dk \nonumber\\
&= x (\int\limits_{0}^{2 \pi} R_k(x) + A_k(x) dk ),
\end{align}
where 
\begin{align*}
R_k(x) &= - 4 \frac{\delta \varepsilon^2_k \Delta_k^2 ( (-x+\delta \varepsilon_k)^2+(x+\delta \varepsilon_k)^2+2\Delta^2_k)}{((-x+\delta \varepsilon_k)^2+\Delta^2_k)^2 ((x+\delta \varepsilon_k)^2+\Delta^2_k)^2} \\
A_k(x) &=  \Delta^2_k \left( \frac{1}{((x+\delta \varepsilon_k)^2+\Delta^2_k)^2} +\frac{1}{((-x+\delta \varepsilon_k)^2+\Delta^2_k)^2}\right) 
\end{align*}
$R_k(x)$ is regular in $k=0$ when $x=0$ ($\lim\limits_{k\rightarrow 0} |R_k(0)| < \infty$) and consequently $|\int\limits_{0}^{2 \pi} R_k(x) |<\infty $  when $x\rightarrow 0$. Conversely, $A_k(0) \propto \frac{1}{k^2}$ when $k\ll 1$. One can then show that $\int\limits_{0}^{2 \pi} A_k(x) dk$ diverges as $\frac{1}{x}$. We finally obtain, as $A_k(x)>0$:
\begin{equation}
\lim\limits_{x\rightarrow 0} \Delta DI(x) = \lim\limits_{x\rightarrow 0} x \int\limits_{0}^{2 \pi} A_k(x) dk > 0
\end{equation}
The derivative is indeed discontinuous at a phase transition where the winding number of $\theta_k$ changes.\\
The proof becomes more involved when $\delta \varepsilon_k$ is also linear in $k$. In that case, both $A_k(x)$ and $R_k(x)$ contributes to the discontinuity, but with opposite signs. At some special, fine-tuned points, the discontinuity may consequently vanish. \\
The proof is straightforwardly extended for several closings of the gap at non-zero momenta.

\section{Structure factor of the bipartite charge fluctuations}
In this section, we demonstrate the properties of the structure factor of the BCF. Let us define the structure factor by:
\begin{align}
SF(A, \phi) =  \langle \sum\limits_{j_1, j_2 = 1}^l Q_{j_1} Q_{j_2} e^{i \phi (j_1-j_2)} \rangle
\end{align}
It is easy to express the structure factor (SF) in a form similar to Eq. \ref{eq:BCF1D-uni}:
\begin{multline}
SF_{\hat{Q}}(A, \phi) = q_e l \iint_{\text{BZ}^2} \frac{dk dq}{16\pi^2} f(k-q, l) \\
\left( 1-\cos(\theta_k) \cos(\theta_{q+\phi})+\sin(\theta_k) \sin(\theta_{q+\phi}) \right),\label{eq:BCF1D-SF}
\end{multline}
The scaling laws of the SF are the same as those of the BCF:
\begin{equation}
SF_{\hat{Q}}(A, \phi) = i_{\hat{Q}}(\phi) l + b(\phi) \log(l) + O(1)
\end{equation}
Both coefficients carry information on the structure of the gap closing.

\subsection{Linear term $i_{\hat{Q}}(\phi)$}
Let consider the SF at a critical point. The linear contribution can be simply obtained from Eq. \ref{eq:BCF1D-SF}:
\begin{multline}
i_{\hat{Q}}(\phi)-\frac{1}{4} = \int\limits_{\text{BZ}} \frac{dk}{4 \pi} \sin(\theta_k) \sin(\theta_{k+\phi})\\
-\int\limits_{\text{BZ}} \frac{dk}{4 \pi} \cos(\theta_k) \cos(\theta_{k+\phi})
\end{multline}
Let $(k_j)_{1\leq j\leq N}$ the momenta at which the gap closes. $\theta_k$ is discontinuous at each $k_j$ such that $\delta_j=\sin(\theta_{k_j^+})-\sin(\theta_{k_j^-}) \neq 0$ (resp. $\delta_j'=\cos(\theta_{k_j^+})-\cos(\theta_{k_j^-}) \neq 0$)  as long as $\Delta_k$ vanishes linearly at $k_j$. We assume that it is the case here. Then one can easily express:
\begin{equation}
\partial_\phi i_{\hat{Q}}(\phi) = \sum\limits_{j=1}^N \frac{\sin(\theta_{(k_j-\phi)})}{4 \pi} \delta_j-\frac{\cos(\theta_{(k_j-\phi)})}{4 \pi} \delta_j' + R(\phi),
\end{equation}
where $R(\phi)$ is a continuous function of $\phi$. Now, to study the discontinuity of $\partial_\phi i_{\hat{Q}}$, we introduce $\Delta I(\phi_0, \delta \phi)=\partial_\phi i_{\hat{Q}}(\phi_0+\delta \phi)-\partial_\phi i_{\hat{Q}}(\phi_0-\delta \phi)$. We discard $R$ as its contribution vanishes in the limit $\delta \phi \rightarrow 0$, leading to the simple expression:
\begin{multline}
\Delta I(\phi_0, \delta \phi)=\sum\limits_{j=1}^N \frac{\sin(\theta_{(k_j-\phi_0-\delta\phi)})-\sin(\theta_{(k_j-\phi_0+\delta\phi)})}{4 \pi} \delta_j\\-\frac{\cos(\theta_{(k_j-\phi_0-\delta\phi)})-\cos(\theta_{(k_j-\phi_0+\delta\phi)})}{4 \pi} \delta_j'
\end{multline}
$\Delta I(\phi_0, 0^+)\neq 0$ only if $\theta_k$ is discontinuous at $k_j+\phi_0$, \textit{i.e. } when the gap closes at $k_j+\phi_0$. Note that at some specific, fine-tuned points when there exist several combinations $(j_1, j_2)$ such that $\phi_0=k_{j_1}-k_{j_2}$ or at specific values of $\theta_{k_j}$, the discontinuity may vanish. 

\subsection{Logarithmic term $b(\phi)$}
Let us now focus on the logarithmic coefficient. From Eq. \ref{eq:BCF1D-fourier} and \ref{eq:BCF1D-SF}, it is straightforward to show that a logarithmic term can appear only at a phase transition. The problem can be simplified to the following one: under which conditions on $g$ and $\phi$, a logarithmic term can arise in the scaling laws of:
\begin{equation*}
I_g=l\iint_{\text{BZ}^2} dk dq f(k-q, l) g(k) \overline{g}(k+\phi),
\end{equation*}
where $g$ is continuous by part and periodic on the Brillouin Zone.\\
\underline{One discontinuity:} let us assume that $g$ is only discontinuous at a single momentum $k_f$ and let us fix $g(k_f^+)-g(k_f^-)=\delta g$. Then, the only possible logarithmic contribution to $I_g$ comes from:
\begin{align*}
J_1(l)=\sum\limits_{n=-l}^l |n| \frac{\delta g e^{i n k_f}}{2\pi n} \times \frac{\delta g e^{-i n (k_f-\phi)}}{2\pi n}= \frac{\delta g^2 }{2\pi^2}\sum\limits_{n=1}^l \frac{\cos( \phi n) }{ |n|} 
\end{align*}
As long as $\phi \neq 0 [2 \pi]$, $J_1(l)=\mathcal{O}(1)$. No logarithmic contributions appear except when the SF coincides with the BCF.\\
\underline{Two discontinuities:} let us now assume that $g$ is discontinuous at two different momenta $k_1$ and $k_2$, with amplitude $\delta g_1$ and $\delta g_2$. Now, possible logarithmic contributions to $I_g$  arise from:
\begin{align*}
J_2(l)&=\sum\limits_{n=-l}^l |n| \frac{\delta g_1 e^{i n k_1}+\delta g_2 e^{i n k_2}}{2\pi n}\times\\
&\qquad \qquad \frac{\delta g_1 e^{-i n (k_1-\phi)}+\delta g_2 e^{-i n (k_2-\phi)}}{2\pi n}\\
&= \frac{1 }{2\pi^2}\sum\limits_{n=1}^l (\delta g_1^2 + \delta g_2^2)\frac{\cos( \phi n)}{n}\\
&\qquad  +\delta g_1 \delta g_2 \frac{\cos((k_1-k_2-\phi)n)+\cos((k_1-k_2+\phi)n)}{ |n|} 
\end{align*}
Logarithmic contributions now also appear at $\phi=\pm(k_1-k_2)$, with a different amplitude $\frac{\delta g_1 \delta g_2}{2\pi^2}$.\\
In the models we consider, symmetries such as parity, time-inversion or particle-hole symmetry often enforce $k_1=-k_2$. Taking into accounts all the symmetry constraints on $\varepsilon_k$ and $\Delta_k$, actual contributions to the SF arise from:
\begin{equation}
\frac{q_e }{2\pi^2}\sum\limits_{n=1}^l \frac{\cos(2 \theta_{k_1})\cos( \phi n)}{n}-
  \frac{\cos((2 k_1-\phi)n)+\cos((2k_1+\phi)n)}{ |n|},
\end{equation}
when the gap close at $\pm k_1$. Note that the amplitude of the logarithmic contributions at $\phi \neq 0$ does not depend on $\theta_k$.\\
\underline{More discontinuities:} one can arbitrarily consider more exotic transitions where the gap closes at $N>2$ different momenta. Computations are not significantly affected and logarithmic contributions will appear at all phases $\phi$ that correspond to the difference between two gap-closing momenta. Universality of the coefficient is also recovered.

\begin{figure}
\begin{center}
\includegraphics[width=0.9\linewidth]{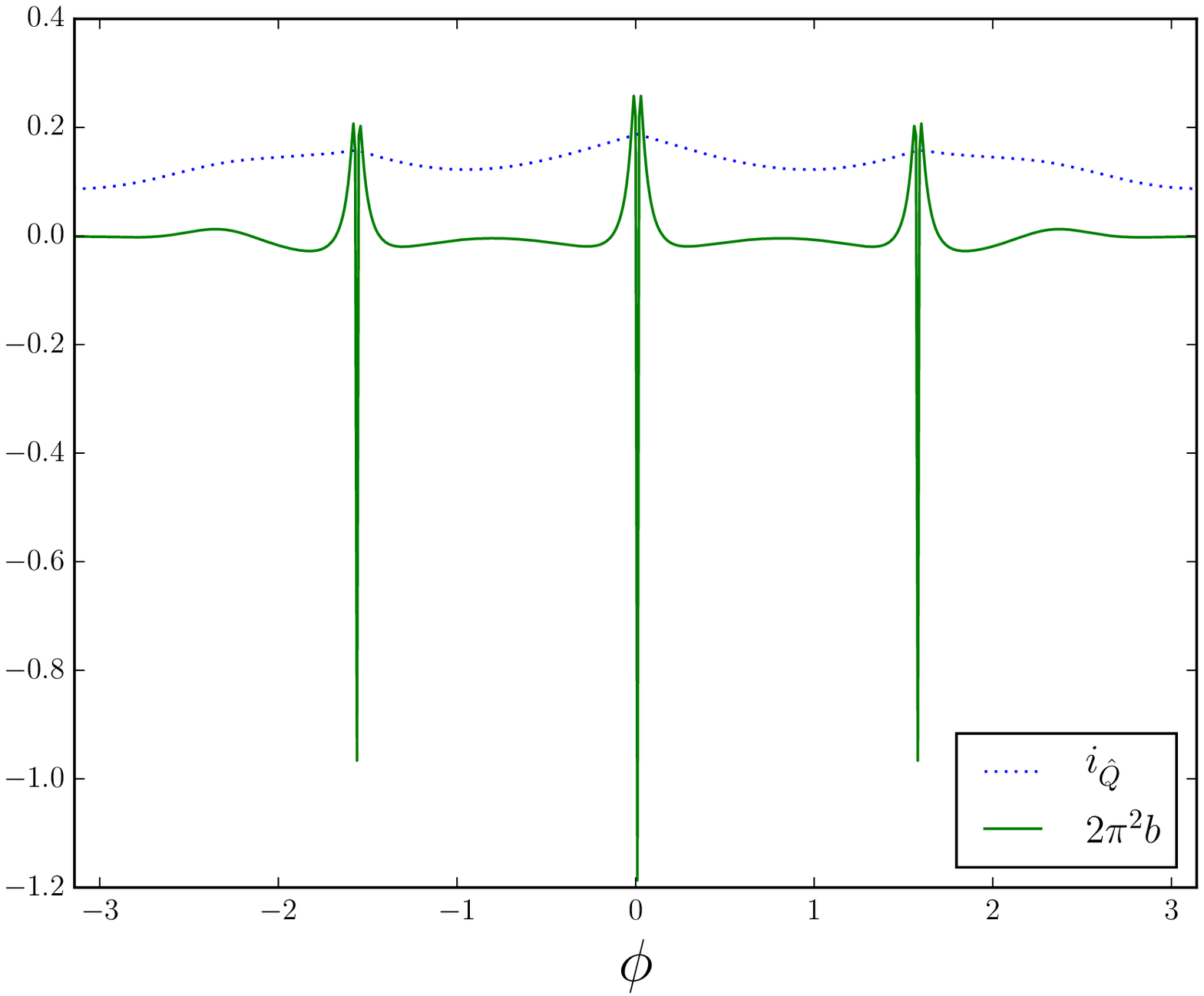}
\end{center}
\caption{Example of QFID and logarithmic contributions to the structure factor obtained from a numerical fit, as a function of $\phi$, in the extended Kitaev model[\hyperref[S1]{S4}] for $t_3=-t$, $\mu=2 \sqrt{2}t$ (on the critical line between $m=0$ and $m=2$. We consider a subsystem of up to $l=1000$ sites. The gap closes linearly at momenta $k_F=\pm 2 \arctan(1+\sqrt{2})$. The logarithmic coefficient is bounded by $\frac{1}{\pi^2}$ but does not saturate ($b(0)\approx -\frac{0.6}{\pi^2}$). To recover the band structure, we vary $\phi$. Logarithmic contributions first vanish then reappear as two sharp peaks at $\pm 2k_F$, and saturate at the universal value $-\frac{1}{2\pi^2}$. The change in the sign of $b$ close to these peaks is an artifact of the fit. We also measure discontinuities of $i_{\hat{Q}}(\phi)$ at $\pm 2k_F$.}
\label{fig:SFactor}
\end{figure}

\section{Bipartite charge fluctuations in the Rashba nanowire}
In this section, we simply expose the main computation steps for diagonalizing the Rashba nanowire model[\hyperref[S1]{S7}, \hyperref[S1]{S8}], and the expression for the various correlation functions that appear in the computation of the BCF.\\
The real-space Hamiltonian for the Rashba model for topological superconductor is (summations on $j$ are implicit):
\begin{multline}
H=-\mu c^\dagger_{j, \sigma} c_{j, \sigma} -t (c^\dagger_{j,\sigma} c_{j+1, \sigma} + h.c.) + V c^\dagger_{j, \alpha} \sigma^z_{\alpha, \beta} c_{j, \beta} \\
-i\lambda (c^\dagger_{j, \alpha} \sigma^y_{\alpha, \beta} c_{j+1, \beta} + h.c.)
+\Delta (c^\dagger_{j, \uparrow} c^\dagger_{j, \downarrow} + c_{j, \downarrow} c_{j, \uparrow}),
\end{multline}
where $c$ are spin-$\frac{1}{2}$ fermionic annihilation operator, $\mu$ is a chemical potential, $t$ a hopping term, $V$ a Zeeman field, $\lambda$ a Rashba spin-orbit coupling and $\Delta$ a $s$-wave pairing obtained by proximity.\\
The model can be exactly diagonalized in the Nambu basis $\Psi_k=(c_{k, \uparrow}, c_{k, \downarrow}, c^\dagger_{-k, \downarrow}, c^\dagger_{-k, \uparrow})^T$. In this basis, the Hamiltonian can be rewritten as:
\begin{equation*}
H=\sum\limits_k \Psi^\dagger_k h(k) \Psi_k
\end{equation*}
with 
\begin{equation}
h(k)=\begin{pmatrix}
\varepsilon(k) + V & -i \varepsilon_2(k) & \Delta & 0\\
i \varepsilon_2(k) & \varepsilon(k)-V & 0 & -\Delta \\
\Delta & 0 & -\varepsilon(k) + V & -i \varepsilon_2(k) \\
0& -\Delta & i \varepsilon_2(k)& -\varepsilon(k)-V
\end{pmatrix},
\end{equation}
with $\varepsilon(k)=-\mu-2t \cos(k)$ and $\varepsilon_2(k)=2 \lambda \sin(k)$.
Defining the Pauli matrices,
\begin{equation*}
\tau_z=\begin{pmatrix}
\mathbb{I}_2 & 0_2 \\
0_2 & -\mathbb{I}_2
\end{pmatrix}, \qquad
\tau^x=\begin{pmatrix}
0_2 & \mathbb{I}_2 \\
\mathbb{I}_2 & 0_2 
\end{pmatrix}, \qquad
\sigma^z=\begin{pmatrix}
 \sigma^z & 0_2 \\
0_2 & \sigma^z 
\end{pmatrix} 
\end{equation*}
the system is diagonalized by:
\begin{multline}
e^{-i \frac{\gamma_+ }{2}R_+} e^{-i \frac{\gamma_- }{2}R_- }e^{i \frac{\beta}{2} \tau^y \sigma^z}e^{-i \frac{\alpha}{2} \sigma^x}h(k) e^{i \frac{\alpha}{2}  \sigma^x} e^{-i \frac{\beta}{2} \tau^y \sigma^z}e^{i \frac{\gamma_+ }{2}R_+} e^{i \frac{\gamma_- }{2}R_- } \\
= \begin{pmatrix}
E_+ & 0 & 0 & 0 \\
0 & E_- & 0 & 0 \\
0 & 0 & -E_- & 0 \\
0 & 0 & 0 & -E_+
\end{pmatrix},
\end{multline}
where the energy spectrum is given by:
\begin{equation}
E^2_\pm = \Delta^2+\varepsilon(k)^2+\varepsilon_2^2(k)+V^2 \pm 2 \sqrt{V^2 \varepsilon(k)^2 +  \varepsilon (k)^2 \varepsilon_2(k)^2+ \Delta^2 V^2},
\end{equation}
the diagonalizing angles by:
\begin{align*}
\alpha &= \text{Arg}(V \varepsilon(k) + i \varepsilon \varepsilon_2)\\
\beta &=  \text{Arg}(\sqrt{V^2 \varepsilon^2+\varepsilon^2 \varepsilon^2_2} + i \Delta V)\\
\gamma_\pm &= \text{Arg}(A_\pm + i B_\pm)\\
A_\pm&=\varepsilon \cos(\beta) \pm \varepsilon_2 \sin(\alpha) + \cos(\alpha) \left(\pm V + \Delta \sin(\beta)\right)\\
B_\pm&=\varepsilon_2 \cos(\alpha) \sin(\beta) - \sin(\alpha) \left(\pm \Delta + V \sin(\beta)\right)
\end{align*}
and the two rotation matrix by:
\begin{align*}
X_+&=\begin{pmatrix}
0 & 0 & 0 & 1 \\
0 & 0 & 0 & 0 \\
0 & 0 & 0 & 0 \\
1 & 0 & 0 & 0
\end{pmatrix}, \qquad 
X_-=\begin{pmatrix}
0 & 0 & 0 & 0 \\
0 & 0 & 1 & 0 \\
0 & 1 & 0 & 0 \\
0 & 0 & 0 & 0
\end{pmatrix}.
\end{align*}
We can express the original fermion in the Bogoliubov basis:
\begin{equation}
\vectq{c_{k, \uparrow}}{c_{k, \downarrow}}{c^\dagger_{-k, \downarrow}}{c^\dagger_{-k, \downarrow}} =e^{i \frac{\alpha}{2}  \sigma^x} e^{-i \frac{\beta}{2} \tau^y \sigma^z}e^{i \frac{\gamma_+ }{2}R_+} e^{i \frac{\gamma_- }{2}R_- } \vectq{\eta_{k, +}}{\eta_{k, -}}{\eta^\dagger_{-k, -}}{\eta^\dagger_{-k, +}},
\end{equation}
such that $H=E_{k,+} \eta^\dagger_{k,+}\eta_{k,+}+E_{k,-}\eta^\dagger_{k,-}\eta_{k,-}$. Computation of the relevant two-fermions correlator in the ground state is straightforward, albeit tedious. 
\begin{multline*}
\Braket{c^\dagger_{k, \uparrow} c_{q, \uparrow}}=\frac{1}{4}\delta_{k,q} \left(2-\cos(\beta) (\cos(\gamma_{+}) + \cos(\gamma_-))\right. \\+\cos(\alpha)\left.(\cos(\gamma_-)-\cos(\gamma_+)) + \sin(\alpha)\sin(\beta)(\sin(\gamma_+)+\sin(\gamma_-)) \right)
\end{multline*}\begin{multline*}
\Braket{c^\dagger_{k, \downarrow} c_{q,  \downarrow}}=\frac{1}{4}\delta_{k,q} \left((2-\cos(\beta) (\cos(\gamma_{+}) + \cos(\gamma_-))\right.  \\+\cos(\alpha)\left.(\cos(\gamma_+)-\cos(\gamma_-)) - \sin(\alpha)\sin(\beta)(\sin(\gamma_+)+\sin(\gamma_-)) \right)
\end{multline*}
 \begin{multline*} 
\Braket{c^\dagger_{k, \uparrow} c^\dagger_{q, \uparrow}}=\frac{-i}{4}\delta_{k,-q} \left(\cos(\beta)(\sin(\gamma_-)+\sin(\gamma_+))\right. \\+\cos(\alpha)\left.(\sin(\gamma_+)+\sin(\gamma_-))+\sin(\alpha) \sin(\beta) ( \cos(\gamma_-)+\cos(\gamma_+))  \right)
\end{multline*}
\begin{multline*}
\Braket{c^\dagger_{k, \downarrow} c^\dagger_{q, \downarrow}}=\frac{-i}{4}\delta_{k,-q}\left(\cos(\beta) (\sin(\gamma_-)+\sin(\gamma_+))\right. \\+\cos(\alpha)\left.(\sin(\gamma_-)-\sin(\gamma_+))-\sin(\alpha)\sin(\beta) (\cos(\gamma_-) +\cos(\gamma_+))\right) 
\end{multline*}

\begin{multline*}
\Braket{c^\dagger_{k, \uparrow} c_{q, \downarrow}} =\frac{-i}{4} \delta_{k,q} \left( \cos(\alpha) \sin(\beta) (\sin(\gamma_-)+\sin(\gamma_+))\right.  \\+ \sin(\alpha) \left.(\cos(\gamma_+)-\cos(\gamma_-)) \right)
\end{multline*}
\begin{multline*}
\Braket{c^\dagger_{k,\uparrow} c^\dagger_{q, \downarrow}}=-\frac{1}{4}\delta_{k,-q} \left( \cos(\alpha) \sin(\beta) (\cos(\gamma_+) + \cos(\gamma_-))   \right)
\end{multline*}
The BCF for the different charges can be safely computed using Wick theorem and the previous expressions. We focus on the topological transition that occur for $-2t<\mu<0$ and $V=\sqrt{\Delta^2+(\mu+2t)^2}$. For large Zeeman field, the system is in a topological phase, while it is a trivial superconductor at low $V$. None of the angles gives a good winding number, but a discontinuity at $k=0$ appear in $\gamma_-$ at the phase transition, marking the topological change.
\begin{figure}
\begin{center}
\includegraphics[width=0.8\linewidth]{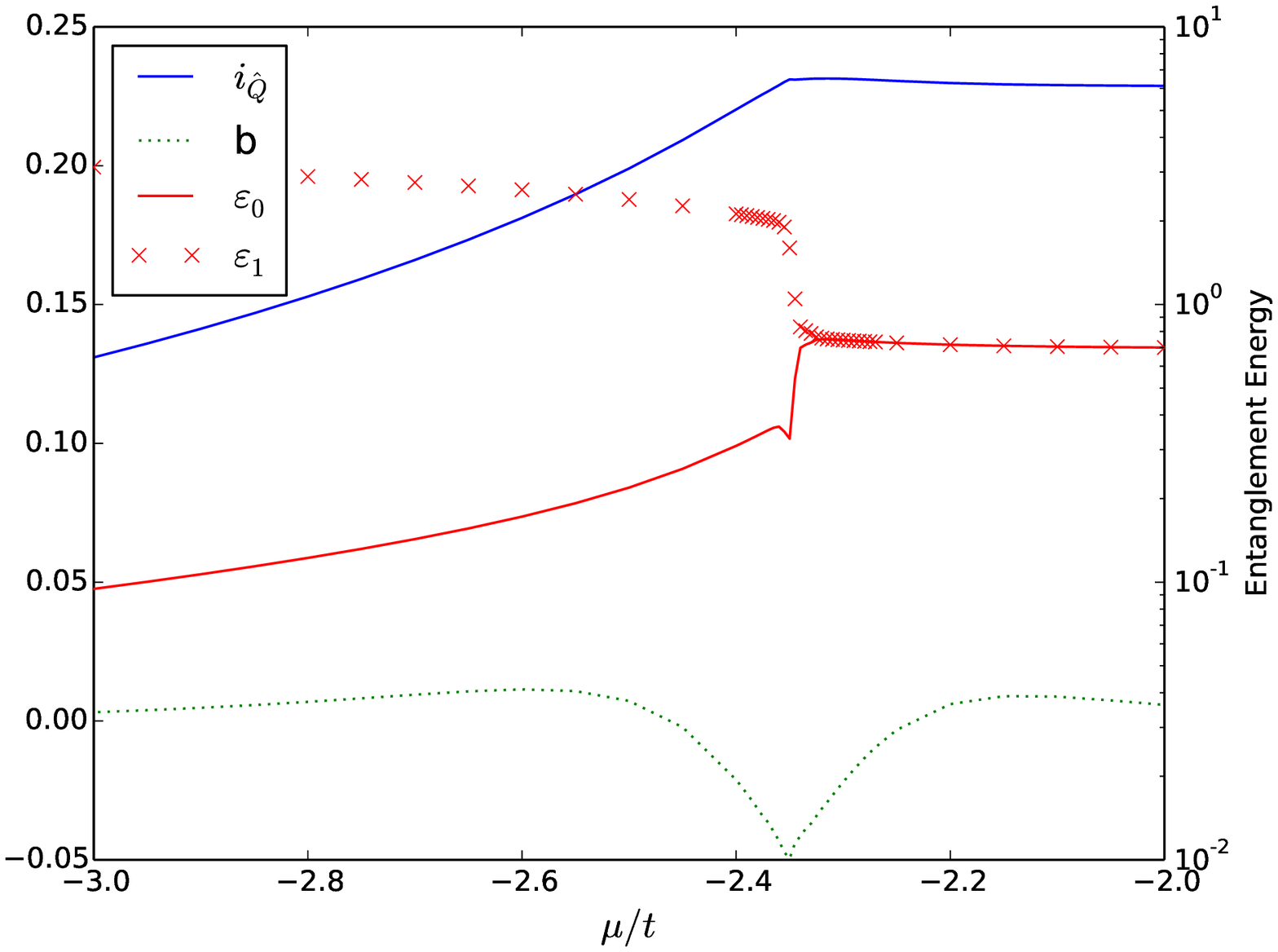}
\end{center}
\caption{Numerical simulations from MPS for a wire of length $L=80$ sites, $\Delta=t$ and $g=0.4t$ with periodic boundary conditions. In red are represented for reference the first two level of the entanglement spectrum. The entanglement spectrum has an exact twofold degeneracy in the topological phase, while it is non-degenerate in the trivial phase. In blue (dotted green), the linear (logarithmic) coefficient obtained from a numerical fit. Note the discontinuity (peak) that still marks the phase transition.}
\label{fig:Interacting}
\end{figure}
\section*{Bipartite charge fluctuations in interacting systems}
We present in this section an example of BCF in an interacting system. The results are obtained from numerical simulations using Alps library for Matrix Product States[\hyperref[S1]{S9}, \hyperref[S1]{S10}]. We consider the an interacting version of Kitaev chain:
\begin{multline}
H=-t(c^\dagger_j c_{j+1} + c^\dagger_{j+1} c_j) + \Delta (c^\dagger_j c^\dagger_{j+1} + c_{j+1} c_j) \\
-\mu c^\dagger_j c_j + g(c^\dagger_j c_{j}-\frac{1}{2})(c^\dagger_{j+1} c_{j+1}-\frac{1}{2}).
\end{multline}
The phase diagram of this model, equivalent to the XYZ spin chain, has been thoroughly investigated[\hyperref[S1]{S11}, \hyperref[S1]{S12}] and we study the phase transition between the topological phase inherited from Kitaev's chain, and a trivial quasi-polarized phase. Figure \ref{fig:Interacting} represents the results from one set of simulations. The BCF and the entanglement spectrum point are in good agreement.\\

\section*{}\label{S1}
\small{
\noindent [S1] A. Kitaev, “Unpaired majorana fermions in quantum wires,” \href{dx.doi.rg/10.1070/1063-7869/44/10S/S29}{Physics Uspekhi 44, 131 (2001)}.\newline 
[S2] H. Guo and S.-Q. Shen, Topological phase in a one-dimensional interacting fermion system, \href{http://link.aps.org/doi/10.1103/PhysRevB.84.195107}{Phys. Rev. B 84, 195107 (2011)}.\newline
[S3] W. P. Su, J. R. Schrieffer, and A. J. Heeger, “Solitons in polyacetylene,” \href{http://link.aps.org/doi/10.1103/PhysRevLett.42.1698}{Phys. Rev. Lett. 42, 1698–1701 (1979)}.\newline
[S4] Y. Niu, S.B. Chung, C.-H. Hsu, I. Mandal, S. Raghu, and S. Chakravarty, “Majorana zero modes in a quantum ising chain with longer-ranged interactions,” \href{http://link.aps.org/doi/10.1103/PhysRevB.85.035110}{Phys. Rev. B 85, 035110 (2012)}.\newline
[S5] S. Tewari and J. D. Sau, “Topological invariants for spin-orbit coupled superconductor nanowires,” \href{http://link.aps.org/doi/10.1103/PhysRevLett.109.150408}{Phys. Rev. Lett. 109, 150408 (2012)}.\newline
[S6] M. Trif and Y. Tserkovnyak, “Resonantly tunable majorana polariton in a microwave cavity,” \href{http://link.aps.org/doi/10.1103/PhysRevLett.109.257002}{Phys. Rev. Lett. 109, 257002 (2012)}.\newline
[S7] Y. Oreg, G. Refael, and F. von Oppen, “Helical liquids and majorana bound states in quantum wires,” \href{http://link.aps.org/doi/10.1103/PhysRevLett.105.177002}{Phys. Rev. Lett. 105, 177002 (2010)}.\newline
[S8] R.M. Lutchyn, J.D. Sau, and S. Das Sarma, “Majorana fermions and a topological phase transition in semiconductor-superconductor heterostructures,” \href{http://link.aps.org/doi/10.1103/PhysRevLett.105.077001}{Phys. Rev. Lett. 105, 077001 (2010)}.\newline
[S9] B. Bauer, L.D. Carr, H.G. Evertz, A. Feiguin, J. Freire, S. Fuchs, L. Gamper, J. Gukelberger, E. Gull, S. Guertler, A. Hehn, R. Igarashi, S.V. Isakov, D. Koop, P.N. Ma, P. Mates, H. Matsuo, O. Parcollet, G. Pawowski, J.D. Picon, L. Pollet, E. Santos, V.W. Scarola, U. Schollwck, C. Silva, B. Surer, S. Todo, S. Trebst, M. Troyer, M.L. Wall, P. Werner, and S. Wessel, “The alps project release 2.0: open source software for strongly correlated systems,” \href{http://stacks.iop.org/1742-5468/2011/i=05/a=P05001}{Journal of Statistical Mechanics: Theory and Experiment 2011, P05001 (2011)}.\newline
[S10] M. Dolfi, B. Bauer, S. Keller, A. Kosenkov, T. Ewart, A. Kantian, T. Giamarchi, and M. Troyer, “Matrix product state applications for the alps project,” \href{dx.doi.org/10.1016/j.cpc.2014.08.019}{Computer Physics Communications 185, 3430–3440 (2014)}.\newline
[S11] E. Sela, A. Altland, and A. Rosch, “Majorana fermions in strongly interacting helical liquids,” \href{http://link.aps.org/doi/10.1103/PhysRevB.84.085114}{Phys. Rev. B 84, 085114 (2011)}.\newline
[S12] F. Hassler and D. Schuricht, “Strongly interacting majorana modes in an array of josephson junctions,” \href{http://stacks.iop.org/1367-2630/14/i=12/a=125018}{New Journal of Physics 14, 125018 (2012)}.}
\end{document}